\documentclass[]{PoS}
\usepackage{epsfig} 
\usepackage{graphicx}
\usepackage{amssymb}
\usepackage{natbib}
\usepackage{times}

\def\XMM{{\em XMM-Newton}}

\def\chan{{\em Chandra}}

\def\igrB{{\em IGR\,J08408-4503}}

\def\xte{{\em XTE\,J1739-302}}

\def\j16479{{\em IGR\,J16479-4514}} 
\def\saxj1818{{\em SAX\,J1818.6-1703}}  
\def\ax{IGR\,J18410-0535}

\title{ \XMM\ observations of the supergiant fast X-ray transients 
XTE\,J1739-302, IGR\,J08408-4503 and IGR\,J18410-0535}

\ShortTitle{E. Bozzo: \XMM\ observations of three SFXTs}

\author{\speaker{E. Bozzo}, C. Ferrigno, J.C. Leyder\\
        ISDC - Science Data Centre for Astrophysics, University of Geneva, Chemin d'Ecogia 16, 1290 Versoix, Switzerland.\\
        E-mail: \email{enrico.bozzo@unige.ch}}

\author{L. Stella, G. Israel, A. Giunta\\
        INAF - Osservatorio Astronomico di Roma, Via Frascati 33, 00044 Rome, Italy.\\
        }
        
\author{M. Falanga\\
        International Space Science Institute (ISSI) Hallerstrasse 6, CH-3012 Bern, Switzerland.\\
        }
        
 \author{S. Campana\\
        INAF - Osservatorio Astronomico di Brera, via Emilio Bianchi 46, I-23807 Merate (LC), Italy.\\
        }

\abstract{We report here on the \XMM\ observations of the three supergiant fast X-ray transients
(SFXT) XTE\,J1739-302, IGR\,J08408-4503, and IGR\,J18410-0535. 
For the latter source we only discuss some preliminary results of our data analysis.  
Some interpretation is provided for the timing and spectral behavior of the three sources  
in terms of the different theoretical models proposed so far to interpret the behavior of the 
SFXTs.}  

\FullConference{8th INTEGRAL Workshop "The Restless Gamma-ray Universe" - Integral 2010,\\
		September 27-30, 2010\\
		Dublin Ireland}

\begin{document}

\section{Introduction} 
\label{sec:intro} 

Supergiant fast X-ray transients are a subclass of supergiant X-ray binaries 
(SgHMXBs) which gained in the past few years a great interest due to their 
peculiar behavior in the X-ray domain \citep[see, e.g.,][]{walter07}. 
The few-hours long outbursts displayed by these sources 
have been so far attributed to the presence of a NS accreting matter from   
the extremely clumpy wind of its supergiant companion 
\citep[these clumps are expected to be a factor 10$^4$-10$^5$ denser than 
the homogeneous stellar wind; see e.g.,][]{walter07}. 
Numerical simulations of supergiant star winds indicate that such high density clumps might 
results from instabilities in the wind \citep{oskinova07}.   
\citet{bozzo08} proposed that the X-ray variability of the SFXT sources   
might be driven by centrifugal and/or magnetic ``gating'' mechanisms, that can halt most 
of the accretion flow during quiescence, and only occasionally permit direct accretion 
onto the NS. The properties of these gating mechanisms depend mainly on the NS magnetic field and spin period. 
At odds with the extremely clumpy wind model, in the gating scenario a transition from the regime in which  
the accretion is (mostly) inhibited to that in which virtually all the captured wind material accretes 
onto the NS requires comparatively small variations in the stellar wind 
velocity and/or density, and can easily give rise to a 
very large dynamic range in X-ray luminosity. 
Yet another model was proposed for IGR\,J11215-5952, so far
the only SFXT which displays regular periodic outbursts; this model 
envisages that the outbursts take place when the NS in its orbit 
crosses a high density equatorial region in the supergiant's wind \citep{sidoli07}.     

The \XMM\ observations that we report on this paper were carried out in order to study 
the quiescent emission of the SFXT sources and gain further insight on the mechanism that 
drives their peculiar X-ray activity. 

\section{ \XMM\ data analysis}
\label{sec:data}

For the present study we used two \XMM\ observations of \xte,\ one \XMM\ observation 
of \igrB,\ and one observation of \ax.\ 

The analysis of the \XMM\ data was carried out by using 
standard techniques and the version 9.0 of the \XMM\ Science Analysis System  
(SAS). We refer the reader to \citet{bozzo10} for more details.

\subsection{XTE\,J1739-302} 
\label{sec:xteresults} 

\XMM\ observed \xte\ twice, on 2008 October 1 (hereafter OBS1) and 
on 2009 March 11 (hereafter OBS2). The Epic-PN camera was operated 
in large window mode in the first case and in small window mode 
during the second observation. After the selection of the good time 
intervals was applied, we obtained a total effective 
exposure time of 32~ks and 24~ks for OBS1 and OBS2, respectively. 
\begin{figure*}
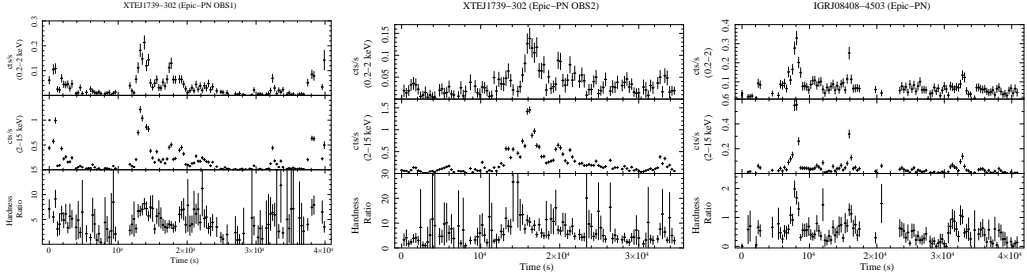

\centering
\includegraphics[scale=0.2,angle=-90]{2008_08_xmm_lc.ps}
\includegraphics[scale=0.2,angle=-90]{2009_xmm_lc.ps}
\includegraphics[scale=0.2,angle=-90]{igrb.ps}
\caption{\XMM\ Epic-PN background-subtracted lightcurves of \xte\ 
during the OBS1 (left panel) and OBS2 (middle panel), and of \igrB\ (right panel). 
In all cases, the upper panel shows 
the source lightcurve in the 0.2-2~keV energy band, whereas the middle panel gives the lightcurve 
in the 2-15~keV energy band (the binning time is 300~s). The ratio of the source count rate in 
the two bands, (2-15~keV)/(0.2-2~keV), versus time is shown in the lower panel.}    
\label{fig:xte2} 
\end{figure*}
From the lower panels in Fig.~\ref{fig:xte2}, 
it is apparent that the source emission hardened  
for higher count rates. A rate resolved analysis 
carried out during the time intervals in which the source count-rate was 
$<0.1$, 0.1-0.4, and $>$0.4 (0.2-15~keV) showed that the hardening of the 
spectrum with the source count-rate was due mostly to a change in the 
cut-off power-law (CUTOFFPL) photon index index \citep[from 1.0 to 1.8; we fixed the cut-off energy at 
13~keV][]{sidoli09}, rather than to variation of the 
absorption column density (consistent with being constant at 
a level of $\sim$2.6$\times$10$^{22}$~cm$^{-2}$). 
The spectra extracted by using the entire exposure time 
of the OBS1 and OBS2, were only poorly fit by a simple absorbed 
CUTOFFPL model. These fits gave 
$\chi^2_{red}$/dof=1.2/175, 1.2/184, for OBS1 and 
OBS2, respectively. We measured in the two cases, 
photon indices of 1.3$\pm$0.1 and 1.19$\pm$0.07 
and absorption column densities $N_{\rm H}$=(2.7$^{+0.2}_{-0.1}$)$\times$10$^{22}$~cm$^{-2}$, 
(3.4$\pm$0.2)$\times$10$^{22}$~cm$^{-2}$. The corresponding 0.5-10~keV X-ray 
fluxes were (2.5$^{+0.3}_{-1.0}$)$\times$10$^{-12}$~erg/cm$^{2}$/s and 
(3.9$^{+0.4}_{-2.0}$)$\times$10$^{-12}$~erg/cm$^{2}$/s. 
\begin{figure}
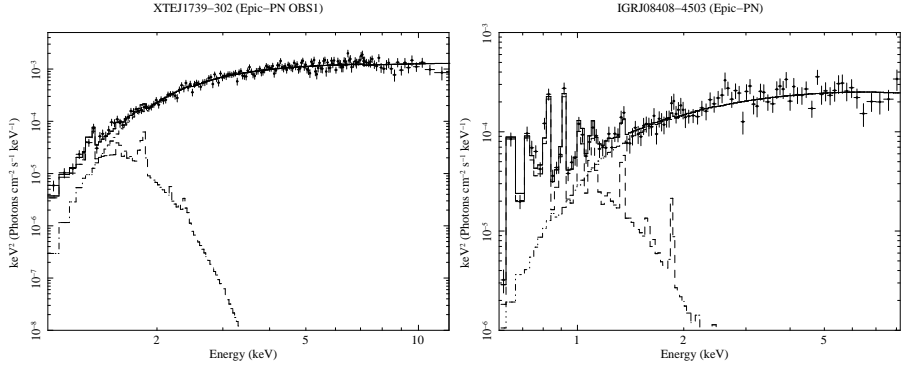

\centering
\includegraphics[scale=0.25,angle=-90]{xtesp1unfolded.ps}
\includegraphics[scale=0.25,angle=-90]{igrbunfolded.ps}
\caption{Unfolded spectra of \xte\ (OBS1, left panel) and 
\igrB\ (right panel). 
The best fit models correspond in both cases to  
an absorbed CUTOFFPL plus a MKL component.}      
\label{fig:xtesp1unfolded} 
\end{figure}
The addition of a thin thermal component ({\sc MEKAL} in {\sc Xspec}; hereafter MKL) 
below 2~keV significantly improved the fits and in the OBS1, where the 
detection of this soft spectral component appeared to be more significant, 
we estimated a temperature of the plasma of 0.15$\pm$0.02~keV and a normalization 
of 0.3$_{-0.2}^{+1.7}$. We checked that a similar fit to the spectrum of the 
OBS2 would give similar results for this spectral component. 
The unfolded spectrum of OBS1 is shown in Fig.~\ref{fig:xtesp1unfolded}.

\subsection{IGR\,J08408-4503} 
\label{sec:igrresults} 

\XMM\ observed \igrB\ on 2007 May 29, with the Epic-PN camera operating in full frame. 
After the selection of the good time intervals was applied, we obtained a total 
effective exposure time of 26~ks. A similar analysis to that described in Sect.~\ref{sec:xteresults} 
performed on the data from this source revealed an analogous behavior to that observed previously in \xte.\ 
From Fig.~\ref{fig:xte2} (left panel), we noticed that the hardness ratio of 
\igrB\ increased with the source count rate. The rate-resolved spectra, 
selected by using the time intervals in which the source count-rate was 
$>$0.2, 0.1-0.2, and $<$0.1, reveled that in this case a simple absorbed CUTOFFPL 
was insufficient to appropriately describe the data. The addition 
of a {\sc MKL} component below $<$2~keV significantly improved these fits. 
In the three spectra, the measured absorption column density was consistent 
(to within the errors) with being constant at a level of 
$\simeq$0.7$\times$10$^ {22}$~cm$^{-2}$, whereas the 
PL photon index showed a significant change decreasing from $\simeq$1.6  
to $\simeq$1.1 at the higher source emission level. The soft spectral component 
did not show any significant variation with the source intensity. We measured 
a temperature of $\simeq$0.2~keV and a normalization of $\simeq$10$^{-3}$. 
Similar values for the spectral parameters were obtained from the fit to the 
spectrum of \igrB\ extracted by using the total exposure time of the \XMM\ observation. 
This spectrum is shown in Fig.~\ref{fig:xtesp1unfolded} (right panel).  

\subsection{IGR\,J18410-0535} 
\label{sec:axresults} 

\begin{figure}
\centering
\includegraphics[scale=0.21]{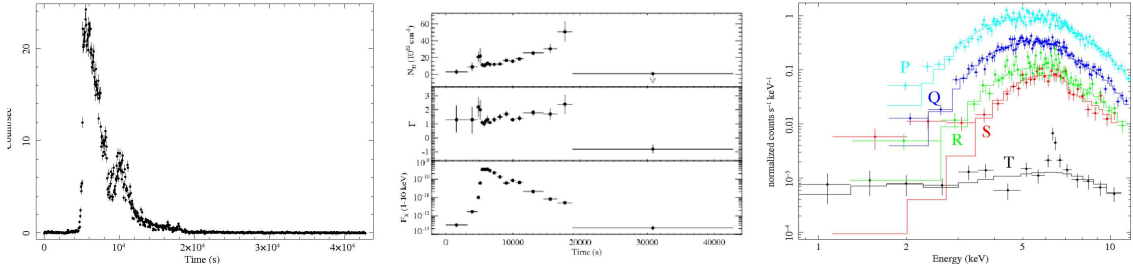}
\caption{Results of the \XMM\ observation of \ax.\ We show in the left 
panel the source lightcurve in the 0.3-12~keV energy band, and report in 
the right panel the results of the time-resolved spectral analysis obtained by dividing the observation 
in 18 time intervals and fitting the corresponding spectra with an absorbed PL model 
($\Gamma$ is the PL photon index, $N_{\rm H}$ the absorption column density and 
$F_{\rm X}$ the observed X-ray flux not corrected for absorption). The right panel show 
the evolution of the source X-ray spectrum during the latest 20~ks of the observation 
(see text for more details).}        
\label{fig:ax} 
\end{figure}

\XMM\ observed \ax\ starting on 2010
March 15 for about 45 ks. The Epic-pn camera was operating
in full frame, while the Epic-MOS1 and Epic-MOS2 cameras
were operating in small window and timing mode, respectively. 
During this observation, IGR J18410-0535 underwent a bright 
X-ray flare that started about 5 ks after the beginning of 
the observation and lasted for $\sim$15~ks. The lightcurve of
the entire \XMM\ observation in the 0.3-12 keV energy
band is shown in Fig.\ref{fig:ax} (left panel). 
Given the high dynamic range in the X-ray flux of the source 
we did not attempt to extract a single spectrum 
by using the total exposure time available. 
Instead we divided the \XMM\ observation in 18 time intervals and perform a preliminary 
fit to the data by using an absorbed power-law (PL) model. The results of these fits 
are reported in Fig.~\ref{fig:ax} (middle panel). 
From this figure we noticed that 
the PL photon index remained fairly constant up to T$\simeq$19000~s, 
whereas the absorption column density underwent dramatic changes. 
Particularly puzzling is the evolution of the source X-ray spectrum during the 
latest 20 ks of observation. This is show in Fig.~\ref{fig:ax} (right panel). 
In this figure the P, Q, R ,S and T spectra are extracted using the time 
intervals T=10420-11520~s, T=11520-14520~s, T=14520-16620~s, T=16620-18870~s, 
and T=18870-42840~s, respectively. 
During the time interval corresponding to the spectrum T, the properties of the 
X-ray emission from \ax\ changed abruptly. The photon index of the PL  
flattened and a prominent emission line at $\sim$6.4~keV clearly emerged from 
the data. A more refined analysis is currently ongoing to understand deeply the nature 
of the spectral changes measured during the X-ray flare. We provide a preliminary 
interpretation of these results in Sec.~\ref{sec:discussion}.

\section{Discussion} 
\label{sec:discussion}

We have presented four deep pointed \XMM\ 
observations of the  
SFXTs \xte,\  \igrB\ and \ax.\  

The first two sources exhibited a very similar behavior. 
Our spectral and timing analysis revealed that both sources 
were characterized by a pronounced variability on time-scales of the 
order of few thousands of seconds. During the \XMM\ observation 
\xte\ and \igrB\ displayed a number of low-intensity flares, 
taking place sporadically from a lower persistent emission level.
The typical duration of these flares is a few thousands of seconds, 
and their X-ray flux is a factor of $\sim$10-30 higher than the persistent 
quiescent flux. Taking into account the fluxes measured during the brightest 
outbursts (See Sect.~\ref{sec:intro}), the total dynamic range
of these two sources is $>$10$^4$. 

The hardness intensity diagrams of \xte\ and \igrB,\ together with the 
results of the rate resolved analysis carried out in Sect.~\ref{sec:data}, 
showed that the variations in the X-ray flux measured during the XMM-Newton 
observations were accompanied in both cases by a change in the spectral 
properties of the sources. In particular we noticed that 
the source hardness ratios were increasing significantly with the X-ray flux. 
A fit to the rate-resolved spectra with a CUTOFFPL model
revealed that this behavior originates from a change in the 
CUTOFFPL photon index, $\Gamma$, rather than in a variation of the 
absorption column density. These results indicate that the timing 
and spectral variability of the two sources during the quiescence 
is qualitatively similar to that observed during the 
outbursts (see Sect.~\ref{sec:intro}). Since most of the X-ray emission 
from these sources is produced by the accretion of matter onto the NS 
\citep{bozzo10}, then the accretion process in these sources
takes place over more than 4 orders of magnitude in the X-ray 
luminosity. 

The properties of the soft component dominating the X-ray spectrum 
of \igrB\ and \xte\ below $\simeq$2~keV could be reasonably
well constrained in the former source, where
the absorption column density was relatively low ($<$10$^{22}$~cm$^{−2}$),
whereas in the case of \xte\ the detection of this component
was less significant. However, the similarity in the timing
and spectral behavior observed in the quiescent state of the two
sources argued in favor of adopting the same spectral model
for both of them. We suggested that a MKL component provide 
a reasonable description of the data, as it would  
represent the contribution to the total X-ray
emission of the shocks in the wind of the supergiant companion.
The results of the fits with this model to the data of the three
observations inferred a temperature of the MKL component and
an emitting region comparable with the values found also in the
case of the SFXT AX\,J1845.0-0433 \citep{zurita09}. 
Similar soft spectral components have been detected in many
other HMXBs and SGXBs. In a few cases, the detection of
 a number of prominent emission lines in the high resolution
X-ray spectra of these sources carried out with the gratings onboard
\chan\ and the RGS onboard \XMM\ \citep[see e.g.,][]{watanabe06} 
have convincingly demonstrated that these
components are produced by the stellar wind around the NS, and
proved to be a powerful diagnostic to probe the structure and
composition of the stellar wind in these systems. The statistics
of the present XMM-Newton observations is far too low to permit
a similar in-depth study of the stellar wind in the case of
\xte\ and \igrB,\ and further longer X-ray observations of these sources  
are probably required in order to firmly establish the nature of the soft 
spectral components in the quiescent emission of the SFXT sources. 

\ax\ underwent a bright X-ray flare during the \XMM\ 
observation, and we reported here some preliminary results of 
the analysis of these data. The time-resolved spectral analysis 
reported in Sect.~\ref{sec:axresults} showed that during this event 
the absorption column density of the X-ray emission from the source 
displayed dramatic changes, whereas the PL photon index 
remained virtually constant up to $T$$\simeq$19000~s. These results suggest 
that the event might have been caused by the accretion of a massive clump, 
similar to the case of IGR\,J17544-2619 \citep{rampy09}. 
A major change in the spectral properties of the source occurred during the latest 20~ks 
of observation, in which our analysis revealed an abrupt flattening of the 
PL photon index and the emergence of a prominent iron-line at 6.4~keV.  
This spectral transition might indicate that at the end of the flare \ax\ underwent some 
kind of obscuration event, similar to what was observed previously in the case 
of IGR\,J16479-4514 \citep{bozzo08b}. A more refined \XMM\ data analysis 
is currently ongoing in order to understand deeply the origin of the 
flare recorded from \ax.\

\end{document}